\documentclass{revtex4-1}
\usepackage{amsmath,amssymb}
\usepackage[pdftex,colorlinks]{hyperref}
\usepackage{graphicx}
\usepackage{color} 
\definecolor{orange}{rgb}{1,0.5,0}

\begin{document}

\title{On the estimators of autocorrelation model parameters}
\author{C. H. Fleming}
\affiliation{Conservation Ecology Center, Smithsonian Conservation Biology Institute, National Zoological Park, Front Royal, VA 22630, USA}
\affiliation{Department of Biology, University of Maryland College Park, MD 20742, USA}
\email{hfleming@umd.edu}
\author{J. M. Calabrese}
\affiliation{Conservation Ecology Center, Smithsonian Conservation Biology Institute, National Zoological Park, Front Royal, VA 22630, USA}
\email{calabresej@si.edu}

\begin{abstract}
Estimation of autocorrelations and spectral densities is of fundamental importance in many fields of science,
from identifying pulsar signals in astronomy to measuring heart beats in medicine.
In circumstances where one is interested in specific autocorrelation functions that do not fit into any simple families of models,
such as auto-regressive moving average (ARMA),
estimating model parameters is generally approached in one of two ways:
by fitting the model autocorrelation function to a non-parameteric autocorrelation estimate via regression analysis
or by fitting the model autocorrelation function directly to the data via maximum likelihood.
Prior literature suggests that variogram regression yields parameter estimates of comparable quality to maximum likelihood.
In this letter we demonstrate that, as sample size is increases, the accuracy of the maximum-likelihood estimates (MLE) ultimately improves by orders of magnitude beyond that of variogram regression.
For relatively continuous and Gaussian processes, this improvement can occur for sample sizes of less than $100$.
Moreover, even where the accuracy of these methods is comparable, the MLE remains almost universally better
and, more critically, variogram regression does not provide reliable confidence intervals.
Inaccurate regression parameter estimates are typically accompanied by underestimated standard errors,
whereas likelihood provides reliable confidence intervals.
\end{abstract}


\maketitle

\section{Introduction}
In all of the physical sciences, and even more generally, one encounters the need to estimate parameters from and identify features of data that are correlated in time or space. 
The most fundamental statistic of the `stochastic process' (time series or spatial field) described by these data is its autocorrelation function or, in the frequency domain, its spectral-density function.
For example, one can identify a pulsar signal, heartbeat or atomic energy-level transition by locating peakedness in the spectral-density function at the relevant frequency.
One can also relate larger spectral features to physical quantities such as temperature, decay rate and `color'.
Often times, in the physical and life sciences, one has a specific model in mind when analyzing data and one desires to estimate the underlying model parameters.
If the given model can be couched in the framework of a particular family of models, such as auto-regressive moving average (ARMA), then parameter estimation may be straightforward.
For ARMA models, the most convenient method of estimating parameters is given by the Yule-Walker equations, whereas the most-reliable and accurate method of estimating parameters is given by maximimum likelihood, \cite{Fan03}.
Unfortunately, many problems of interest cannot be placed within the luxury of frameworks such as ARMA with their well-established estimation proceedures.
For example, given any low-temperature quantum system, all spectral-density functions will take the form
\begin{align}
s(\omega) &= \left[ \hbar \omega \, \coth\!\left( \frac{\hbar \omega}{2 k_\mathrm{B} T} \right) - \hbar \omega \right] g(\omega) \, , \label{eq:quantum}
\end{align}
in accord with the fluctuation-dissipation relation (FDR), \cite{QOS},
where $\hbar$ is Plank's constant, $k_\mathrm{B}$ is Boltzmann's constant, $T$ is the temperature, $\omega$ is the angular frequency and $g(\omega)$ is an even function specified by the system
that may contain additional temperature dependence if the system or operator under consideration is nonlinear.
Such a spectral density cannot be characterized by any finite-parameter ARMA model, due to the presence of the hyperbolic cotangent function.
ARMA models correspond to linear difference equations driven by white noise, and their spectral-density functions take the form of Fourier-Pad\'{e} approximants.
This form of spectral density is not always convenient for describing the processes one would find in physics or biology.

Resorting to general methods, one has two choices for model parameter estimation given stationary processes.
The most convenient method is to first calculate a non-parametric estimate of the autocorrelation function (or spectral density) and then to fit the autocorrelation function to this estimate, as if it were data.
A more rigorous method is to fit the model directly to the data via maximum likelihood.
The prior geostatistics literature indicates that these methods (variogram regression and likelihood) yield comparable accuracy for both real and simulated data, \cite{Zimmerman91,Cressie93},
and the current geostatistics literature continues to refer back to these older comparisons, \cite{Banerjee03,Cressie11}.
However, we found that, when applied to animal-relocation data, variogram regression yielded standard-error estimates that were suspiciously small.
Questions of parameter accuracy (beyond what is reasonable) and questions of confidence-interval accuracy are difficult to answer with real data, necessitating the simulation approach we take here;
while enhanced performance with simulated data does not necessarily carry over to real data, the failure of a method to perform adequately with ideal, simulated data cannot be dismissed.

Following a large number of simulations with a variety of autocorrelation structures and sample sizes, only a small number of which we report here,
we find that these two methods of parameter estimation have comparable point-estimate accuracy only in a limited range of conditions, and outside of this range, maximum likelihood can be vastly superior.
Specifically, the two methods perform similarly, in terms of point estimation, given either non-Gaussian behavior or discontinuity in the underlying processes, accompanied by a small sample size.
To be fair, this regime is of interest in geostatistics, where most of the work comparing these methods has been done, \cite{Mardia84,Warnes87,Zimmerman91,Cressie93}.
However, outside of geostatistics, problems characterized by more continuous, normally-distributed processes and larger sample sizes will be much more common, and the accuracy of maximum likelihood (ML) estimation in such situations often greatly exceeds that of the regression-based approach.
Moreover and more crucially, even in the regime where both methods produce similar point estimates, we find that variogram regression yields unreliable error estimates and confidence intervals when correlation of the errors is not accounted for,
and that this effect, counterintuitively, grows increasingly worse with sample size.
An important consequence of this behavior is that variogram regression can produce heavily-biased parameter estimates accompanied by tight confidence intervals, giving practitioners a false sense of security. 
While it is well known that maximum likelihood performs well for Gaussian random variables,
we have found no clear demonstration of the conditions under which it is outperformed by alternative, standard methods when the estimation of model autocorrelation functions is the goal.
Despite the common knowledge of maximum likelihood's performance and capability, there is widespread use of inferior techniques in estimating the parameters of model autocorrelation functions that do not conform to specific time-series models like ARMA, \cite{Cressie93}.

\section{Methods}
\subsection{Least-squares and $\chi^2$ regression analysis}

Currently dominant methods of parameter estimation for autocorrelation functions start by first estimating the autocorrelation non-parametrically via the (semi-) variogram $\hat{\gamma}(t)$, covariogram or periodogram $\hat{s}(f)$.
For instance, the unbiased method-of-moments estimate of the semi-variogram is given by
\begin{align}
\hat{\gamma}(t) &= \frac{1}{2 \, n(t)} \sum_{t'} \left[ x(t'\!+\!t) - x(t') \right]^2 \, , \label{eq:variogram}
\end{align}
where $n(t)$ denotes the number of data pairs with lag $t$.
This estimate is then treated as data and the sum of squared residual between the autocorrelation estimate and autocorrelation function of interest, $\gamma(t|\theta)$, is minimized with respect to the parameters $\theta$
\begin{align}
L(\theta) &= \sum_t w(t) \left[ \hat{\gamma}(t)-\gamma(t|\theta) \right]^2 , \label{eq:LS}
\end{align}
and with weights $w(t)$.
While it is true that both the variogram and the periodogram are more general than any specific form of likelihood function that will necessarily make distributional assumptions,
this generality is lost in the second stage of calculation when the autocorrelation function parameters are estimated.
The least-squares method makes distributional assumptions in that the residuals are taken to be approximately Gaussian and uncorrelated.
By the Gauss-Markov theorem, this method is equivalent to an uncorrelated Gaussian likelihood function for $\hat{\gamma}(t)$ with an unknown variance proportional to $1/w(t)$.
One of the most convenient and reasonable choices of weighting is to use the $\chi^2$ variance of $\gamma^2/n$, \cite{Cressie85}.
A fully $\chi^2$ fit of the variogram can easily be calculated with the same $\mathcal{O}(n)$ computational cost,
however, there are larger issues with this analysis, such as the lack of account for correlation in $\gamma$, between different lags.
The consideration of correlated errors is possible within the least-squares framework,
however, it is both easier and more consistent (in terms of distributional assumptions) to use likelihood for this task.

The one exception of a convenient approach that is also rigorous and accounts for the correlation of errors is a $\chi^2$ (or similar) fit to the periodogram.
In this case, it is known that such a method is asymptotically equivalent to the exact likelihood function we consider,
with asymptotic $\mathcal{O}(1/n)$ errors, \cite{Whittle53,Dembo86,Gray06}.
However, the regime of validity for this method is smaller in scope:
The data must be evenly sampled and, in higher dimensions, assembled in a rectangular grid.
There are certain situations, such as when the spectrum varies greatly in amplitude, where round-off errors in the Fourier transform will contaminate the result.
Finally, as was mentioned, this method is an asymptotic approximation and so sample sizes must be sufficiently large so that the $\mathcal{O}(1/n)$ errors are negligible.
If these conditions are met, then all of our statements pertaining to the quality and reliability of ML carry over to $\chi^2$ fitting the periodogram.
The relevant log-likelihood function for discrete Fourier transformed (DFT) data is given by
\begin{align}
\ell(\theta|x) &= \sum_f \left[ -\frac{1}{2} \log s(f|\theta) - \frac{\left| \tilde{x}(f) - \tilde{\mu}(f|\theta) \right|^2 dt}{2 \, s(f|\theta)} \right] , \label{eq:Whittle} \\
\tilde{x}(f) &= \frac{1}{\sqrt{n}} \sum_t e^{+2\pi \imath f t} \, x(t) \, ,
\end{align}
where the range of frequencies is given by $f \in \{0,df,\cdots,(n\!-\!1)df\}$ (with the latter half being effectively negative),
the frequency resolution is given by $df = 1/n \, dt$, in terms of the temporal resolution $dt$,
and the model mean is given by $\mu(t)$, which need not be stationary.
Likelihood function \eqref{eq:Whittle} is simply Whittle's approximate likelihood function,
but without ignoring the normalization of the spectral density and without assuming a stationary mean.

\subsection{Likelihood analysis}
\begin{figure}
\begin{center}
\includegraphics[width=5in]{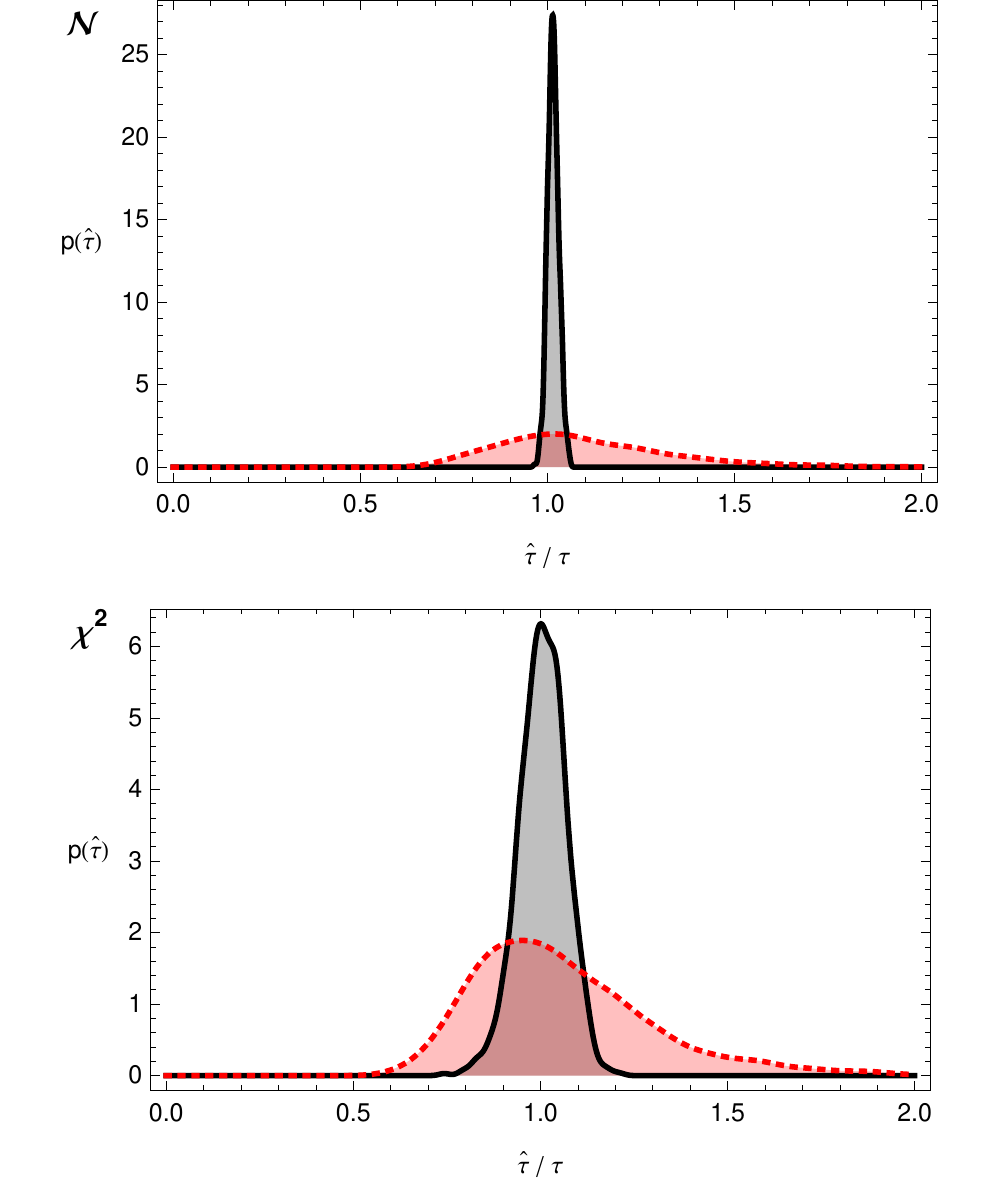}
\end{center}
\caption{\textbf{Distributional effect on accuracy}:
parameter-estimate distributions for normal $\boldsymbol{\mathcal{N}}$ and $\boldsymbol{\chi^2}$ noise,
both having Gaussian-shaped autocorrelation functions with correlation time $\tau$, a sampling frequency of $1/2\tau$ and a sample period of $32 \, \tau$.
The smoothed histogram of 1000 $\tau$ estimates are compared for maximum-likelihood \textcolor{black}{$\boldsymbol{-}$} and variogram regression \textcolor{red}{$\boldsymbol{\cdots}$}.
The accuracy advantage in the likelihood estimator is diminished, but still present, as the distributional assumptions are violated.}
\label{fig:gaussian}
\end{figure}

As discussed, when fitting a model to an autocorrelation estimate, some distributional assumptions must be made about the data.
The most widely accepted methods all take the data to be Gaussian in estimating autocorrelation errors,
whereas the autocorrelation function itself might be also taken to be Gaussian -- at least for the convenience of a least-squares regression.
Obviously, it is more consistent to simply use a Gaussian distribution for the data,
and we will argue that this choice is generally suitable even if the data are not normally distributed.
There are several ways to see its general applicability:
\begin{enumerate}
\item When considering unstructured random variables (or processes), such a likelihood function reproduces the standard formulas for the mean and covariance, \cite{Pawitan01},
which are valid for a wide variety of distributions.
\item Such a likelihood function reproduces the standard formulas for the periodogram, \cite{Dembo86},
and the Fourier transform makes absolutely no assumption that a process is Gaussian.
\item It is known that such an estimator is asymptotically normal and unbiased for any causal and invertible ARMA model,
regardless of the distribution of the underlying white noise, \cite{Fan03}.
Such asymptotic normality therefore extends to a much larger class of autocorrelation functions, assuming their spectral density is subject to Fourier-Pad\'{e} approximation.
This may be the case even when ARMA methods are not practical, as with quantum correlations \eqref{eq:quantum}.
\item For a given mean and covariance, the distribution of maximum entropy is the Gaussian distribution, \cite{Cover06}.
Therefore according to the principle of maximum entropy, the Gaussian assumption the most suitable in the absence of any information regarding the higher-order cumulants.
\end{enumerate}
In Figure~\ref{fig:gaussian},
we compare different methods of parameter estimation for Gaussian and $\chi^2$ processes with the same autocorrelation structure.
For highly non-Gaussian processes, the accuracy advantage of the MLE is diminished, and the orders-of-magnitude improvement is staved off to larger-and-larger sample sizes.
However, at no point does the MLE perform worse than the variogram regression, which makes similar distributional assumptions.
Furthermore, the lack of reliability in the confidence-interval estimation of the variogram regression remains an issue, as we shall discuss.

The exact log-likelihood function is given by
\begin{align}
\ell(\theta) &= - \frac{1}{2} \log \det (\boldsymbol{\Sigma}(\theta)) - \frac{1}{2} \left( \mathbf{X} - \mathbf{M}(\theta) \right)^\mathrm{T} \boldsymbol{\Sigma}(\theta)^{-1} \left( \mathbf{X} - \mathbf{M}(\theta) \right) \, , \label{eq:L}
\end{align}
to within a constant,
where the data vector, mean vector, and autocorrelation matrix is given by
\begin{align}
X_i &= x(t_i) \, , \\
M_i &= \left\langle x(t_i) \right\rangle , \\
\Sigma_{ij} &= \left\langle x(t_i) \, x(t_j) \right\rangle ,
\end{align}
where $\langle \cdots \rangle$ denotes the arithmetic average.
Some general methods for maximizing the likelihood of parametric models are given by \cite{Smith01}.

\subsection{Analysis}
\label{sec:detail}
The likelihood analysis often yields an orders-of-magnitude increase in parameter estimation accuracy for sufficiently large sample sizes.
As we will show, for suitably continuous and Gaussian processes, the relevant sample size for which this improvement occurs can be less than $100$.
However, as the underlying process becomes increasingly discontinuous and non-Gaussian, the orders-of-magnitude effect occurs only for larger-and-larger sample sizes,
though the overall efficiency of ML remains higher than that of the regression-based approach.
To isolate the effect of continuity, we consider the class of spectral-density functions
\begin{align}
s(f) &\propto \frac{1}{\left[ 1 + (2\pi f\tau)^2 \right]^{1+d}} \, , \label{eq:rational}
\end{align}
where $\tau$ determines the relevant correlation timescale and $d$ is an integer that determines the degree of continuity for the underlying process.
E.g,. for $d=0$ the model is Ornstein-Uhlenbeck, which has a process that is continuous but not smooth;
for $d=1$ the process is smooth but has discontinuous curvature, etc..
This class of spectral densities is approximately described by ARMA$(d,0)$ when $dt \ll \tau$,
but it is restricted to only one timescale parameter $\tau$, instead of $d$ timescale parameters.

To generate our correlated Gaussian processes, we subsampled from larger data sets generated by the FFT algorithm of \cite{Billah90}.
This algorithm is considerably faster (for large data sets) and more stable than the conventional method of transforming white noise with matrices.
In this way, time-series data can be easily simulated with sample sizes for which the likelihood calculations can barely be managed. 
To investigate the effect of non-Gaussian noise we generated Gaussian noise and performed power-law transformations.
For instance, in Figure~\ref{fig:gaussian} we squared the time series, transforming it from Gaussian to $\chi^2$.
Then using the moment-decomposition formulas for a multivariate Gaussian process, one can relate the moments of the Gaussian process to the moments of the $\chi^2$ process.

\section{Results}
\label{sec:parametric}

We subjected each method of analysis to a wide variety of autocorrelation functions and conditions, the most critical of which we report here.
We found a vast improvement in the estimation of autocorrelation function shape parameters when maximizing the likelihood function, e.g. correlation time, decay rate, temperature, etc..
For evenly-sampled data sets, the ordinary statistical estimates of the mean and variance (that neglect autocorrelation structure) were closer to the MLE.
Variogram-regression estimates of the variance were also closer to the MLE, however their corresponding standard-error estimates were unreliable.
There does appear be an effect whereby the sampling distributions of the MLEs of the mean and variance improve over the regression estimates,
but this occurs far more gradually with increasing sample size as compared to the shape parameters.
This much smaller increase in efficiency was noticed by \cite{Zimmerman91}, 
however, the larger improvement in shape-parameter estimation was not noticed there for the Ornstein-Uhlenbeck process.

\begin{figure}
\begin{center}
\includegraphics[width=5in]{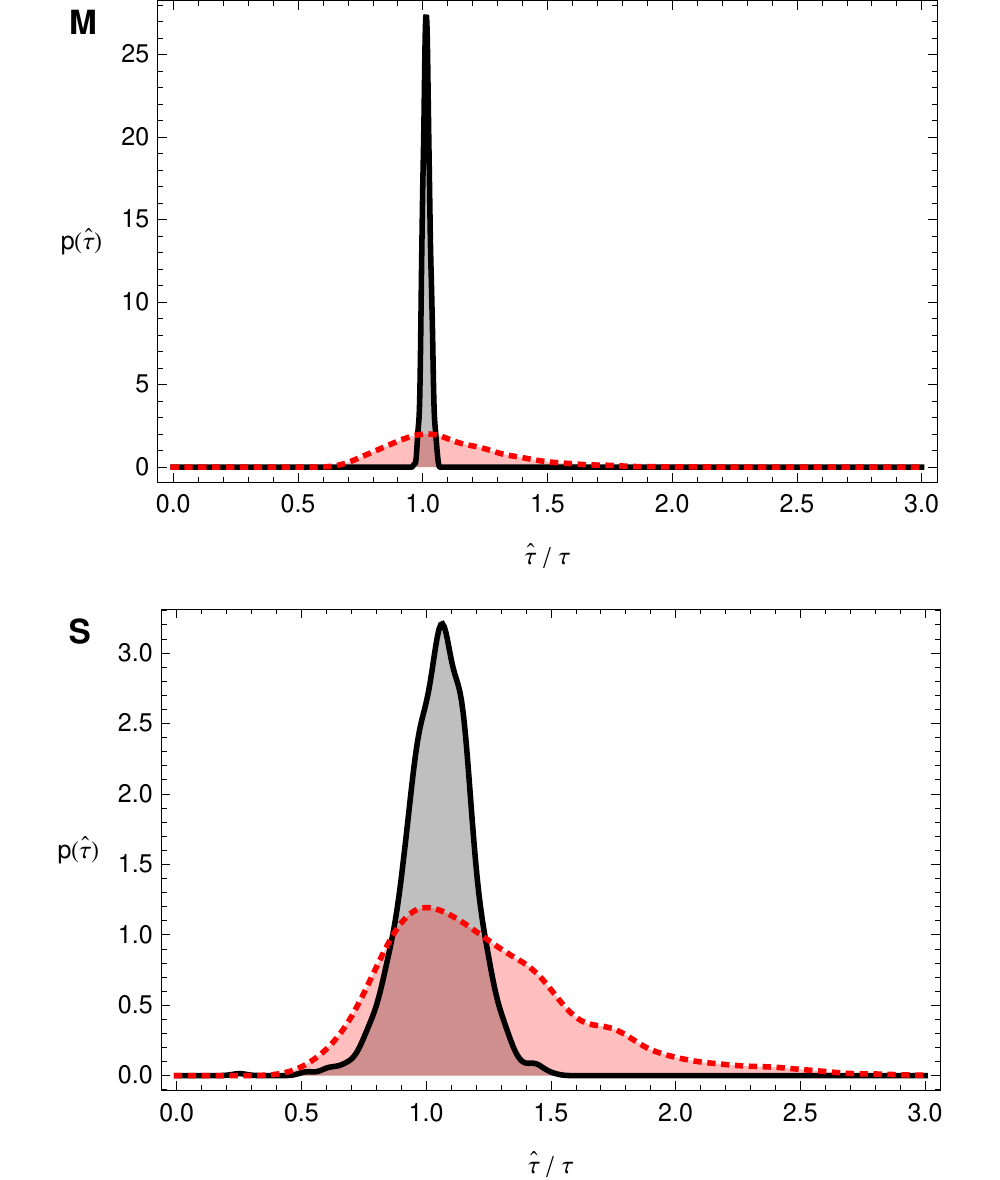}
\end{center}
\caption{\textbf{Sample-size effect on accuracy}:
parameter-estimate distributions for Gaussian-shaped autocorrelation functions with correlation time $\tau$.
The \textbf{M}edium sample has a sampling frequency of $1/2\tau$ and a sample period of $32 \, \tau$,
while the \textbf{S}mall sample has a sampling frequency of $1/\tau$ and a sample period of $16 \, \tau$.
The smoothed histogram of 1000 $\tau$ estimates are compared for maximum-likelihood \textcolor{black}{$\boldsymbol{-}$} and variogram regression \textcolor{red}{$\boldsymbol{\cdots}$}.
The likelihood estimate undergoes rapid improvement with sample size as compared to the regression estimate.}
\label{fig:size}
\end{figure}

In Figure~\ref{fig:size},
we demonstrate how the orders-of-magnitude increase in parameter-estimation accuracy of the MLE develops with larger sample sizes.
In general, the parameter estimation is universally better, however,
such a vast difference in the accuracy of the parameter estimates does not present itself until the sample size becomes sufficiently large.
In this case, the process under consideration in Figure~\ref{fig:size} is infinitely continuous and Gaussian,
and the orders-of-magnitude improvement can be achieved with a sample size of less than $100$.
As the data becomes increasingly non-Gaussian or discontinuous, 
the overall efficiency remains higher, though the rate at which the accuracy of the MLEs pulls away with increasing sample size is diminished,
increasing the sample size at which an orders-of-magnitude improvement would be observed.
We also observed a qualitative difference in the point-estimate sampling distributions.
For the MLE, this sampling distribution was very localized around the true value of the parameters,
whereas the sampling distribution of the regression estimate had long tails,
indicating a much higher sampling variance (i.e., lower efficiency) on the same datasets relative to maximum likelihood.

In Figure~\ref{fig:gaussian},
we demonstrate how large violations of the distributional assumptions diminish, but do not eliminate, the greater parameter-estimation accuracy of maximum likelihood. 
The $\chi^2$ noise considered in Figure~\ref{fig:gaussian} is particularly non-Gaussian, as the $\chi^2$ distribution is relatively skewed and has a lower limit of zero (i.e. is non-negative).
It is worth reiterating that both the likelihood and the variogram regression approaches make similar distributional assumptions about the data.
Therefore the intent of such an investigation into non-Gaussian processes is to determine if the likelihood method somehow falls behind variogram regression (it does not).
There are several reasons that likelihood remains more accurate.
First, the likelihood method only treats the data Gaussian, whereas the variogram regression treats the variogram as both Gaussian and $\chi^2$ for different purposes, with only the latter being consistent with the data being treated as Gaussian.
Regardless of the distributional properties of the data, the variogram is cannot become more Gaussian than $\chi^2$, as it is always non-negative with a lower limit of (at minimum) zero.
Second, the convenience of variogram analysis results from ignoring error correlation,
whereas the likelihood method appropriately accounts for all correlations in the data to which it is fit, up to the second moment, regardless of its distribution.
This is also a strict improvement that is not lost as the data becomes increasingly non-Gaussian.

\begin{figure}
\begin{center}
\includegraphics[width=5in]{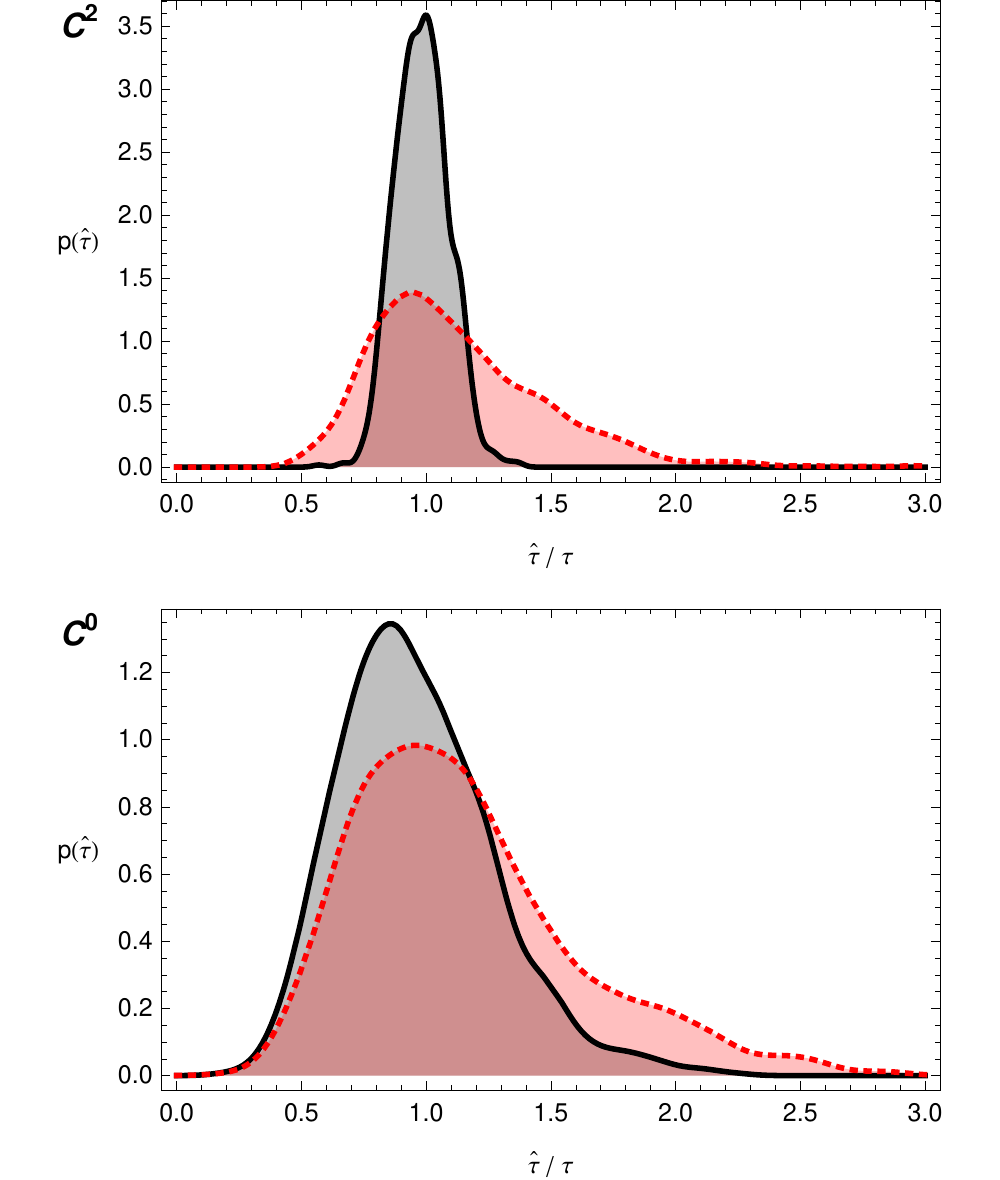}
\end{center}
\caption{\textbf{Continuity effect on accuracy}:
parameter-estimate distributions for model \eqref{eq:rational} with correlation time $\tau$ and varying degree $d=0,2$ of continuity $C^d$,
given a sampling frequency of $1/2\tau$ and a sample period of $32 \, \tau$.
The smoothed histogram of 1000 $\tau$ estimates are compared for maximum-likelihood \textcolor{black}{$\boldsymbol{-}$} and variogram regression \textcolor{red}{$\boldsymbol{\cdots}$}.
In terms of sample size, it is easier to achieve a marked improvement in parameter estimation with continuous processes.}
\label{fig:continuity}
\end{figure}

As demonstrated by Figure~\ref{fig:continuity},
increasing the degree of discontinuity in the underlying process also diminishes, but does not eliminate, the greater parameter-estimation accuracy of maximum likelihood. 
Whereas our other comparisons consider infinitely-continuous processes,
the two processes in Figure~\ref{fig:continuity} both have a finite degree of continuity.
This effect explains the findings of comparative analyses in the geostatistics literature and we will expand upon this in the discussion.

\begin{figure}
\begin{center}
\includegraphics[width=5in]{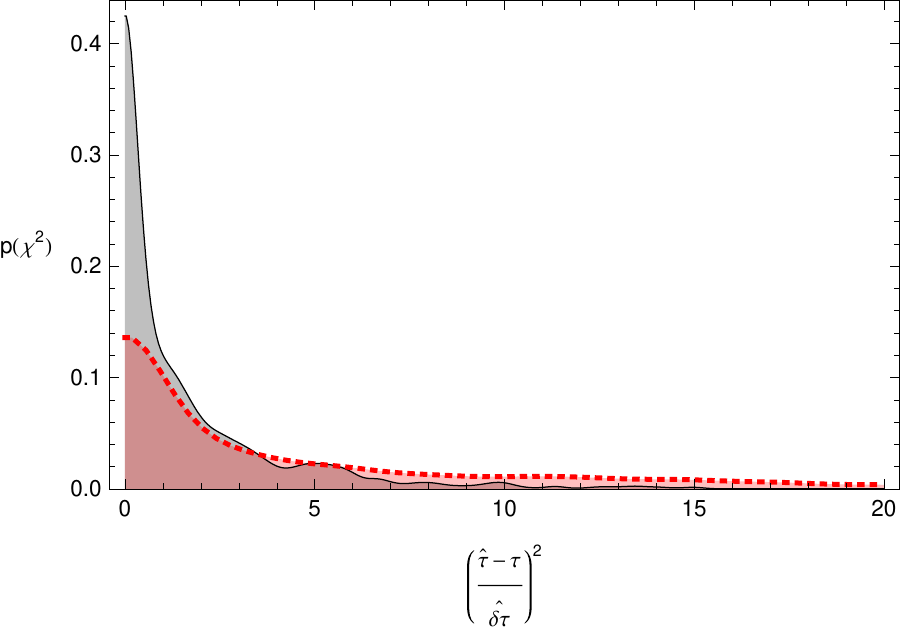}
\end{center}
\caption{\textbf{Reliability of confidence intervals}: The distribution of (squared) parameter errors relative to the estimated errors,
from the \textbf{M}edium sample in Figure~\ref{fig:size},
for maximum likelihood $\boldsymbol{-}$ and variogram regression \textcolor{red}{$\boldsymbol{\cdots}$}.
The variogram distribution here is extremely long tailed even though its parameter distribution is not.
The regression estimate suffers from a severe underestimation of standard errors.
}
\label{fig:residual}
\end{figure}

\begin{figure}
\begin{center}
\includegraphics[width=5in]{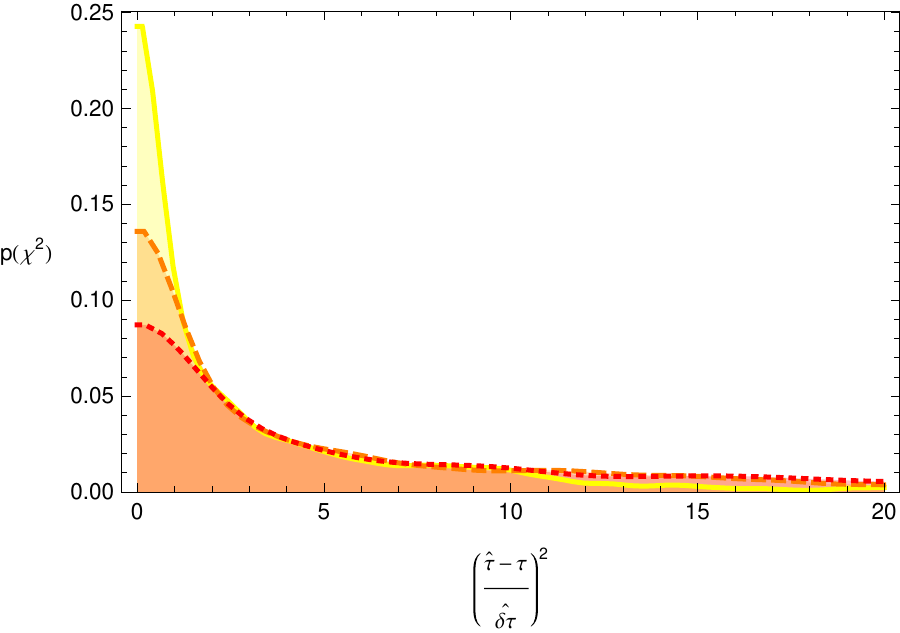}
\end{center}
\caption{\textbf{Limiting behavior of variogram-derived standard errors}:
from the \textbf{S}mall \textcolor{yellow}{$\boldsymbol{-}$} and \textbf{M}edium \textcolor{orange}{\textbf{- -}} samples of Figure~\ref{fig:size},
and a larger sample \textcolor{red}{$\boldsymbol{\cdots}$}, with further doubled period and frequency.
The parameter-estimate distributions are converging with increasing sample size, albeit much more slowly than with ML,
however the performance of the standard error estimate becomes increasingly poor.}
\label{fig:badlimit}
\end{figure}

One of our most troubling findings, which does not seem to have been considered in the literature,
regards the behavior of the confidence-interval estimates.
Even when the sample size is large enough so that the distribution of parameter estimates by variogram regression is reasonable,
their corresponding confidence-interval estimates are extremely poor.
In Figure~\ref{fig:residual}
we compare the true parameter errors relative to their estimated standard errors for the two methods.
In this regime there is sufficient data for the MLE to obtain an orders-of-magnitude improvement in parameter estimation,
and yet the variogram regression still does not produce reasonable confidence-interval estimates.
In this case, almost half of the regression parameter estimates are more than three (estimated) standard errors from the true value of the parameter,
and it is common for the standard error to be smaller than the true error by a factor of tens.
One would assume that this behavior would improve with increasing sample size, but, in fact, it grows worse, as demonstrated in Figure~\ref{fig:badlimit},
even though parameter accuracy is improving.

\section{Discussion}
Estimating the autocorrelation function of a stochastic process is a fundamental, though difficult, task that is common across the sciences;
often times one has a specific physical or biological model in mind for the mean and autocorrelation functions to fit to.
Applying least-squares regression to non-parametric autocorrelation estimates is a ubiquitous practice,
partly due to the ease with which this can be done and partly due to comparative point-estimate analyses performed in the geostatistics literature where,
as we have shown, conditions of continuity and sample size considered reduce the advantage of ML-based parameter estimation relative to variogram regression.

Here, we have demonstrated the clear superiority of a likelihood-based approach to estimating autocorrelation function parameters.
We have shown that dramatic improvements in the quality of both parameter estimates and interval estimates are often achieved with maximum likelihood.
For larger data sets, maximum likelihood can yield an orders-of-magnitude improvement in parameter estimation accuracy, which can be realized for sample sizes of less than $100$.
General improvements in parameter estimation also hold for smaller data sets and the likelihood method continues to perform reliably for sample sizes so small that regression-based methods break down.
We have also compared confidence-interval estimation between these methods, which to our knowledge, had not yet been done.
We find that variogram regression produces unreliable confidence intervals and, surprisingly, these interval estimates become worse with more data, even though the point estimates improve.
In contrast, maximum likelihood produces reliable confidence intervals across a broad range of conditions, autocorrelation functions, and sample sizes.

It is clear from our results that the likelihood method should be preferred when computationally feasible.
Unfortunately, each evaluation of the likelihood function is computationally expensive, being $\mathcal{O}(n^3)$ in general.
For large and evenly-spaced data sets, it is known that the Whittle approximation to the likelihood function ($\chi^2$ fitting to the DFT periodogram) is equivalent to within $\mathcal{O}(1/n)$ asymptotic errors and,
with the FFT, it has a mere $\mathcal{O}(n \log n)$ computational cost.
When the approximation is valid, the performance advantages in terms of point and interval estimates we have demonstrated for the exact likelihood function carry over to this approximate likelihood function.

One of the most subtle effects discovered in this work is the role of continuity on autocorrelation parameter estimation.
The potential to obtain good parameter estimates is limited by how continuous the underlying stochastic process is,
which manifests statistically in how rapidly the spectral-density function vanishes at high frequencies.
This effect is extremely important in geostatistics,
where the most common autocorrelation behaviors (and all those considered in \cite{Zimmerman91}) are either linear at lag zero,
which indicates a lack of smoothness in the process,
or they have a `nugget effect' at lag zero,
which indicates discontinuity in the process.
However, in many physical and biological situations, stochastic processes can have a considerable degree of continuity when sampled at sufficient frequency.
To give a broad kinematic example, Figure~\ref{fig:continuity}-$C^0$ describes a Brownian motion that is continuous in its trajectory but has divergent velocities,
whereas sub-Figure $C^2$ describes a Brownian motion with finite velocities and finite accelerations but divergent jerks.
This demonstrates that for animal or molecular motion that appears smooth and continuous when sampled with sufficient frequency,
maximum likelihood offers significant benefits.

In geostatistics, the autocorrelation function itself is often not an object of interest, but an intermediate used for interpolating and forecasting.
Otherwise, the increasingly poor performance of variogram regression in determining confidence intervals would have likely been noticed.
We offer two speculations as to why this effect occurs.
First, variogram errors are highly correlated (a property that does not become asymptotically negligible) and the convenient regression-based approach ignores these correlations.
If the variogram estimate is larger or smaller than the true value at one lag,
then it will also tend to be larger or smaller at adjacent lags.
Not taking this correlation of errors into account biases any regression that minimizes a fit residual.
Second, most any log-likelihood function for the data $x$, and not merely the Gaussian likelihood function we consider,
will scale linearly with the data, $\ell_\mathrm{x} \sim n$, and therefore the standard errors will scale according to $\delta \theta_x \sim 1/\sqrt{n}$.
However, a weighted fit to the variogram results in $\ell_\mathrm{\gamma} \sim n^2$ and $\delta \theta_\gamma \sim 1/n$,
as there are geometrically more degrees of freedom in the variogram, when treating it as data, than in the actual data itself.

Least-squares regression of autocorrelation and spectral-density estimates is a convenient and widespread practice in the sciences.
However, as we have demonstrated, there are significant advantages provided by the likelihood approach, 
and we conclude by outlining several situations where it should be seriously considered:
(i) when one requires not only a reliable parameter estimate, but reliable confidence intervals,
(ii) where potential models are known and one seeks the most accurate parameter estimates,
(iii) if the data are evenly sampled, and so there is no extra computational burden in performing a $\chi^2$ fit to the DFT periodogram.

\bibliography{bib}

\end{document}